\documentclass[conference,twocolumns]{IEEEtran}
\usepackage{amsmath,amssymb,amsthm}
\usepackage{graphicx}
\usepackage{psfrag}
\usepackage{stfloats}

\usepackage{mathptmx}

\DeclareMathAlphabet{\mathcal}{OMS}{cmsy}{m}{n}
\SetSymbolFont{operators}{bold}{OT1}{txr}{bx}{n}
\SetSymbolFont{letters}{bold}{OML}{txmi}{bx}{it}
\SetSymbolFont{symbols}{bold}{OMS}{txsy}{bx}{n}
\SetSymbolFont{largesymbols}{bold}{OMX}{txex}{bx}{n}

\DeclareSymbolFont{lettersA}{U}{txmia}{m}{it}
\SetSymbolFont{lettersA}{bold}{U}{txmia}{bx}{it}
\DeclareFontSubstitution{U}{txmia}{m}{it}
\DeclareMathSymbol{v}{\mathalpha}{lettersA}{51}

\makeatletter
\newcommand{\vast}{\bBigg@{3}}
\newcommand{\Vast}{\bBigg@{4}}
\makeatother

\usepackage{bm}

\allowdisplaybreaks[3]

\title{Central Approximation in Statistical Physics and Information Theory}

\author{
\IEEEauthorblockN{Ryuhei Mori and Toshiyuki Tanaka}
\IEEEauthorblockA{
Graduate School of Informatics,
Kyoto University\\
Kyoto, 606--8501, Japan\\
Email: rmori@sys.i.kyoto-u.ac.jp, tt@i.kyoto-u.ac.jp
}
}

\theoremstyle{plain}
\newtheorem{theorem}{Theorem}
\newtheorem{lemma}[theorem]{Lemma}

\begin{document}
\maketitle
\begin{abstract}
In statistical physics and information theory, although the exponent of the partition function is often of our primary interest, 
there are cases where one needs more detailed information. 
In this paper, we present a general framework 
to study more precise asymptotic behaviors of the partition function, 
using the central approximation in conjunction with the method of types. 
\end{abstract}

\section{Introduction}
In information theory and statistical physics, we often face the problem of analyzing a sum of the form
\begin{equation*}
Z = \sum_{\bm{x}\in\mathcal{X}^N} A(\bm{x})
\end{equation*}
where $A(\bm{x})$ is a non-negative function of $\bm{x}\in\mathcal{X}^N$ which is often factorized to local contributions of $\bm{x}$.
In statistical physics, $Z$ is called a partition function.
We also deal with this quantity in information theory~\cite{mezard2009ipa}, \cite{tanaka2002statistical}.
In both of statistical physics and information theory, we are mainly interested in the exponent of $Z$, i.e.,
\begin{equation}\label{eq:fe}
Z = \mathrm{e}^{NF+o(N)}
\end{equation}
or equivalently
$F:=\lim_{N\to\infty}(1/N)\log Z$.
In order to obtain the exponent, while statistical physicists have proposed various techniques,
we have found that the method of types provides a general and intuitive approach~\cite{mori2011connection}.
For obtaining the exponent $F$ in this approach, Laplace's method is used after the classification of $\mathcal{X}^N$ according to the types of assignments.
In this paper, we consider a more precise analysis of $Z$ which is exact up to a constant factor, i.e.,
\begin{equation}\label{eq:pf}
Z = \mathrm{e}^{NF}C(N)(1+o(1))
\end{equation}
where $C(N)$ is a subexponential function of $N$.
Derivation of $C(N)$ is the main purpose of this paper.

In statistical physics and information theory, most of works are dedicated to the analysis of exponent~\eqref{eq:fe}.
Two of the important exceptions are the error probability of the random codes below the critical rate~\cite{gallager1973random}, which uses
a finer version of Cram{\'e}r's theorem~\cite{bahadur1960deviations},
and
the expected number of codewords of low-density parity-check (LDPC) codes~\cite{1715529}.
In both cases, the {\it central approximation}~\cite{flajolet2009analytic} is essentially used.
This paper shows the usefulness of the central approximation combined with the method of types for many models in statistical physics and information theory.

\section{Notations and Useful equations}
Let $A(a,b)$ be the $(a,b)$ element of a matrix $A$.
Let $I_n$ be the identity matrix of size $n$.
Let $A^t$ be the transpose of $A$.
Let $|\mathcal{X}|$ be the cardinality of a set $\mathcal{X}$.
Let $Dg$ and $D^2g$ be the gradient and the Hessian matrix of a function $g$.
Let $\|\bm{z}\|$ be the $L_2$ norm of $\bm{z}$.
Let $\mathcal{H}(\nu)$ denote the entropy function $-\sum_{x\in\mathcal{X}}\nu(x)\log \nu(x)$ of a probability measure $\nu$ on $\mathcal{X}$.
The following lemmas are used in this paper.
\begin{lemma}[Sylvester's determinant theorem]\label{lem:sylvester}
For $n\times m$ matrix A and $m\times n$ matrix B,
\begin{equation*}
\det(I_n + AB) = \det(I_m + BA).
\end{equation*}
\end{lemma}
\begin{lemma}[Local approximation]\label{lem:local}
For a probability measure $\nu(x)$ on $\mathcal{X}$ satisfying $\nu(x)>0$ for all $x\in\mathcal{X}$
 and a function $v(x)$ satisfying $\sum_{x\in\mathcal{X}} v(x) = 0$,
\begin{align*}
&\binom{N}{\{N\nu(x) + \sqrt{N}v(x)\}_{x\in\mathcal{X}}}=
\frac{\sqrt{2\pi N}}{\prod_{x\in\mathcal{X}} \sqrt{2\pi N\nu(x)}}
\exp\left\{N\mathcal{H}(\nu)\right\}\\
&\quad\cdot \exp\left\{-\sum_{x\in\mathcal{X}}\sqrt{N}v(x)\log \nu(x)
-\sum_{x\in\mathcal{X}}\frac{v(x)^2}{\nu(x)}
\right\}\\
&\quad\cdot \left(1 +\sum_{x\in\mathcal{X}}O\left(\frac{v(x)^3}{\sqrt{N}\nu(x)^2}\right)\right).
\end{align*}
\end{lemma}

\section{Dense model}
\subsection{Asymptotic analysis}
Let $\mathcal{X}\subsetneq\mathbb{R}$ be a finite set.
In this section, without specifying any details, we study 
the following generic ``partition function,'' 
which has the form of a randomness-averaged $n$th power of 
a partition function of a certain dense model: 
\begin{align}
&\mathbb{E}[Z^n] := \sum_{\bm{x}\in\left(\mathcal{X}^n\right)^N} \exp\vast\{\sum_{i=1}^Nf\left(\left\{x_i^{(a)}\right\}_{a\in\{1,\dotsc,n\}}\right)\nonumber\\
&\;+ Ng\left(\left\{\frac1{N}\sum_{i=1}^Nx_{i}^{(a)}x_{i}^{(b)}\right\}_{a\in\{1,\dotsc,n\},\,b\in\{a,\dotsc,n\}}\right)\vast\}
\label{eq:dense}
\end{align}
where $f$ and $g$ are bounded continuous functions taking $n$ and $n(n+1)/2$ arguments, respectively.
The function $g$ is assumed to be invariant under permutations of replica indices $a,\,b$, and to have a Hessian matrix.
This model includes as special cases various models 
often studied in statistical physics and information theory,
e.g., the Sherrington-Kirkpatrick (SK) model~\cite{mezard2009ipa}, random matrices~\cite{edwards1976eigenvalue}, code-division multiple-access channels~\cite{tanaka2002statistical}, etc.
By using the method of types, one obtains~\cite{monasson1998optimization}
\begin{align*}
&\mathbb{E}[Z^n] = \sum_{v(\bm{x})} \binom{N}{\{v(\bm{x})\}}\exp\vast\{\sum_{\bm{x}\in\mathcal{X}^n}v(\bm{x})f\left(\left\{x^{(a)}\right\}_{a\in\{1,\dotsc,n\}}\right)\\
&\;+ Ng\left(\left\{\frac1{N}\sum_{\bm{x}\in\mathcal{X}^n}v(\bm{x})x^{(a)}x^{(b)}\right\}_{a\in\{1,\dotsc,n\},\,b\in\{a,\dotsc,n\}}\right)\vast\}
\end{align*}
where $v(\bm{x})$ is a type of length $N$ on the alphabet $\mathcal{X}^n$. 
From Laplace's method, 
the exponent $F:=\lim_{N\to\infty}(1/N)\log\mathbb{E}[Z^n]$ is given by 
\begin{align}
&F
= \max_{\nu(\bm{x})}\vast\{ \mathcal{H}(\nu)
+\left\langle f\left(\{x^{(a)}\}_{a\in\{1,\dotsc,n\}}\right)\right\rangle_\nu\nonumber\\
&\quad + g\left(\left\{\left\langle x^{(a)}x^{(b)}\right\rangle_\nu\right\}_{a\in\{1,\dotsc,n\},\,b\in\{a,\dotsc,n\}}\right)\vast\}
\label{eq:lap}
\end{align}
where $\nu(\bm{x})$ denotes a probability measure on $\mathcal{X}^n$,
and where
\begin{equation*}
\langle a(\bm{x})\rangle_\nu:=\sum_{\bm{x}\in\mathcal{X}^n}\nu(\bm{x})a(\bm{x})
\end{equation*}
for any function $a(\bm{x})$.
Here, we consider a more detailed result on $\mathbb{E}[Z^n]$ of the form~\eqref{eq:pf}.
In fact, the factor $C(N)$ in this case does not depend on $N$,
and is obtained via the central approximation~\cite{flajolet2009analytic}.
\begin{theorem}[Central approximation for the dense model]\label{thm:dense}
Assume that the solution of the maximization problem~\eqref{eq:lap} is unique and is denoted by $\nu^*(\bm{x})$.
Furthermore, assume $\nu^*(\bm{x})>0$ for all $\bm{x}\in\mathcal{X}^n$
and
\begin{equation*}
\det\left(I_{n(n+1)/2} - D^2g(U'-U)\right)>0
\end{equation*}
where $U'$ and $U$ are $n(n+1)/2\times n(n+1)/2$ matrices defined by
\begin{align*}
U'((a,b),(c,d))&= \langle x^{(a)}x^{(b)}x^{(c)}x^{(d)}\rangle_{\nu^*}\\
U((a,b),(c,d))&= \langle x^{(a)}x^{(b)}\rangle_{\nu^*}\langle x^{(c)}x^{(d)}\rangle_{\nu^*}.
\end{align*}
Then,
\begin{equation*}
\mathbb{E}[Z^n] = \mathrm{e}^{NF}
\det\left(I_{n(n+1)/2} - D^2g(U'-U)\right)^{-\frac12}(1+o(1))
\end{equation*}
where $F$ is given by \eqref{eq:lap}.
\end{theorem}
Note that if the solution of the maximization problem~\eqref{eq:lap} is not unique,
the constant factor is
\begin{equation*}
\sum_{\nu^*(\bm{x})}\det\left(I_{n(n+1)/2} - D^2g(U'-U)\right)^{-\frac12}
\end{equation*}
where the contributions from all solutions $\nu^*(\bm{x})$ of the maximization problem~\eqref{eq:lap} are summed up.
For the $p$-spin model~\cite{mezard2009ipa}, $D^2 g$ is a diagonal matrix whose diagonal elements are
\begin{align*}
D^2g((a,b),(a,b))&=
\beta^2\binom{p}{2}\langle x^{(a)}x^{(b)}\rangle_{\nu^*}^{p-2}
\end{align*}
where $\beta>0$ is inverse temperature.
The positive definiteness of the matrix for which the determinant is taken is equivalent to the Almedia-Thouless (AT) condition~\cite{almeida1978stability},
which is a condition for the stability of a replica symmetric (RS) solution.

\subsection{On the replica symmetric assumption}
In the replica theory, we often assume the RS assumption, i.e., $\nu^*(\bm{x})$ is invariant under permutations of the $n$ variables in $\bm{x}$.
In this section, for simplicity, it is assumed that the alphabet is $\mathcal{X}=\{+1,-1\}$.
The matrices $D^2g$, $U'$ and $U$ can thus be reduced to $n(n-1)/2\times n(n-1)/2$ matrices since $x^{(a)}x^{(a)}=1$ always holds.
It is known that $D^2g$ and $U'-U$ share the same eigenspaces~\cite{almeida1978stability}, \cite{tanaka2002statistical}.
Let $A$ be the $n(n-1)/2\times n(n-1)/2$ matrix with elements
\begin{equation}
A((a,b),(c,d))=\begin{cases}
P,& \text{ if } |\{a,b\}\cap\{c,d\}|=2\\
Q,& \text{ if } |\{a,b\}\cap\{c,d\}|=1\\
R,& \text{ if } |\{a,b\}\cap\{c,d\}|=0.
\end{cases}
\label{eq:PQR}
\end{equation}
Both $U'-U$ and $D^2g$ are of this form on the RS assumption.
The eigenvectors of $A$ does not depend on $P$, $Q$ and $R$.
From this observation, one obtains
\begin{align*}
&\det\left(I_{n(n-1)/2}-D^2g(U'-U)\right) =\\
&\bigg(1
-\left(1-q^2 + 2(n-2)q(1-q) + \frac{(n-2)(n-3)}2 (r-q^2)\right)\\
&\qquad\cdot\left(P+2(n-2)Q+\frac{(n-2)(n-3)}2R\right)\bigg)\\
&\cdot\Big(1-\left(1-q^2 + (n-4)q(1-q) - (n-3)(r-q^2)\right)\\
&\qquad\cdot\left(P+(n-4)Q-(n-3)R\right)\Big)^{n-1}\\
&\cdot\left(1-\left(1-q^2 - 2q(1-q) + r-q^2\right)\left(P-2Q+R\right)\right)^{\frac{n(n-3)}2}
\end{align*}
where $P$, $Q$ and $R$ are \eqref{eq:PQR} for $D^2g$ and where
\begin{align*}
q&:=\langle x^{(a)}x^{(b)}\rangle_{\nu^*},&
r&:=\langle x^{(a)}x^{(b)}x^{(c)}x^{(d)}\rangle_{\nu^*}.
\end{align*}
In the definitions of $q$ and $r$, the indices $a$, $b$, $c$ and $d$ are all different.
At the limit $n\to0$, the finite-size correction term of the RS free energy $\mathbb{E}[\log Z]/N$ is
\begin{align*}
&\lim_{n\to 0}\frac1n\frac1N\log\det\left(I_{n(n-1)/2}-D^2g(U'-U)\right)^{-\frac12} \\
&=-\frac1{2N}\bigg[\log\left(1-\left(1- 4q +3r\right)\left(P-4Q+3R\right)\right)\\
&\quad -\frac32\log\left(1-\left(1-2q+r\right)\left(P-2Q+R\right)\right)\bigg]
\end{align*}
where the variables $q$, $r$, $P$, $Q$ and $R$ are to be determined by the saddle point condition of the RS free energy~\cite{mezard2009ipa}.
For the SK model where $P=\beta^2$, $Q=R=0$, in the paramagnetic phase $\beta<1$ where $q=r=0$, the finite-size correction term is
$(1/(4N))\log\left(1-\beta^2\right)$.
This result is known in~\cite{parisi1993critical}.
For the SK model, at the critical temperature $\beta=1$, eigenvalues of the Hessian include zero.
For $\beta > 1$ where the full-step replica symmetry breaking must be considered, the Hessian also includes zero eigenvalue. 
Hence, for $\beta\ge 1$, the second derivative analysis is not sufficient and the analysis of third or higher-order derivative is needed~\cite{flajolet2009analytic}.
For $\beta= 1$, the results are partially obtained in~\cite{parisi1993critical}.

\subsection{Proof}
The proof of Theorem~\ref{thm:dense} is the same as the ordinary proof using the saddle point method~\cite{flajolet2009analytic}.
Let $\alpha\in(1/2,2/3)$.
The equations for deriving Theorem~\ref{thm:dense} are in the next page.
The asymptotic equality $A\sim B$ means that $A=B(1+o(1))$.
From continuity of $f$ and $g$, the assumption of unique maximum, and $\alpha>1/2$, the sum for $\|v(\bm{x})-N\nu^*(\bm{x})\|>N^\alpha$
is asymptotically negligible~\cite{flajolet2009analytic}.
From $\alpha<2/3$, the approximation in Lemma~\ref{lem:local} and
the second-order expansion of $g$ are used
in~\eqref{eq:p2} and \eqref{eq:p3}, respectively.
In~\eqref{eq:p4}, the first-order factor is removed from the optimality of $\{\nu^*(\bm{x})\}$.
In~\eqref{eq:p5}, the Riemann integral formula is used.
In the next equality, the Gaussian integral is performed.
Here, $H$ is the $|\mathcal{X}^n|\times(|\mathcal{X}^n|-1)$ matrix defined by
\begin{equation*}
H(\bm{x},\bm{x}')=\begin{cases}
-1,& \text{if } \bm{x}=\bm{x}_0\\
1,& \text{if } \bm{x}=\bm{x}'\\
0,& \text{otherwise}
\end{cases}
\hspace{2em} \text{for } \bm{x}\in\mathcal{X}^n,\; \bm{x}'\in\mathcal{X}^n\setminus \bm{x}_0
\end{equation*}
for any fixed $\bm{x}_0\in\mathcal{X}^n$, 
$B$ is the $|\mathcal{X}^n|\times|\mathcal{X}^n|$ diagonal matrix defined by $B(\bm{x},\bm{x})=1/\nu^*(\bm{x})$, 
and $J$ is the $|\mathcal{X}^n|\times n(n+1)/2$ matrix defined by $J(\bm{x},(a,b))= x^{(a)}x^{(b)}$.
One obtains Theorem~\ref{thm:dense}
by using Sylvester's determinant theorem (Lemma~\ref{lem:sylvester}) 
and the following equations, which can be verified easily
\begin{align*}
H(H^tBH)^{-1}H^t &= S' - S, &
J^t(S'-S)J&= U' - U
\end{align*}
where $S'=B^{-1}$ and
$S(\bm{x},\bm{x}')=\nu^*(\bm{x})\nu^*(\bm{x}')$.

\subsection{Perturbation of the joint empirical distribution from the i.i.d. Boltzmann distributions}
For $\bm{x}\in(\mathcal{X}^m)^N$, let the $m$-joint empirical distribution be
\begin{equation*}
\nu_m^{\bm{x}}(\bm{z}) := \frac1N\sum_{i=1}^N \prod_{a=1}^m\mathbb{I}\left\{x_i^{(a)}=z^{(a)}\right\}, \hspace{2em} \text{for }\bm{z}\in\mathcal{X}^m.
\end{equation*}
For a Boltzmann distribution with an energy function $E(\bm{x})$, the probability distribution of the joint empirical distribution is defined as
\begin{align*}
P_E(\{\nu(\bm{z})\}) &:=\sum_{\bm{x}\in(\mathcal{X}^m)^N}\frac{\prod_{a=1}^m\exp\{-E(\bm{x}^{(a)})\}}{Z^m}\\
&\quad\cdot \prod_{\bm{z}\in\mathcal{X}^n}\mathbb{I}\left\{\nu_m^{\bm{x}}(\bm{z})\le \nu(\bm{z})\right\}.
\end{align*}
Here, we consider randomness of the energy function and the expectation of $P_E(\{\nu(\bm{z})\})$ with respect to it, i.e., $P(\{\nu(\bm{z})\}):=
\mathbb{E}[P_E(\{\nu(\bm{z})\})]$.
By the replica method, it can be calculated as~\cite{mezard2009ipa}
\begin{align*}
P(\{\nu(\bm{z})\}) &=\lim_{n\to 0}\sum_{\bm{x}\in(\mathcal{X}^n)^N}\mathbb{E}\left[\prod_{a=1}^n\exp\{-E(\bm{x}^{(a)})\}\right]\nonumber\\
&\quad\cdot \frac1{\binom{n}{m}}\sum_{\substack{\mathcal{A}\subseteq \{1,\dotsc,n\},\\|\mathcal{A}|=m}}
\prod_{\bm{z}\in\mathcal{X}^n}\mathbb{I}\left\{\nu_m^{\bm{x}^{(\mathcal{A})}}(\bm{z})\le \nu(\bm{z})\right\}.
\end{align*}
Almost the same calculation as that of $\mathbb{E}[Z^n]$ shows that 
it tends to the delta distribution on the RS assumption~\cite{mezard2009ipa}
\begin{equation*}
\lim_{N\to\infty} P(\{\nu(\bm{z})\}) = \prod_{\bm{z}\in\mathcal{X}^n}\mathbb{I}\left\{\nu_m^{\mathrm{RS}}(\bm{z})\le \nu(\bm{z})\right\}
\end{equation*}
where 
$\nu_m^{\mathrm{RS}}(\bm{x})$ is the $m$-joint distribution determined from the RS solution.
For the dense model, i.e., $\mathbb{E}[Z^n]$ is of the form of~\eqref{eq:dense},
by the same calculation as that of $\mathbb{E}[Z^n]$, a scaled distribution can be obtained from
\begin{align*}
P'(\{\epsilon(\bm{z})\}) &:=\lim_{n\to 0}\sum_{\bm{x}\in(\mathcal{X}^n)^N}\mathbb{E}\left[\prod_{a=1}^n\exp\{-E(\bm{x}^{(a)})\}\right]\nonumber\\
&\;\cdot \prod_{\bm{z}\in\mathcal{X}^n}\mathbb{I}\left\{\sqrt{N}(\nu_m^{\bm{x}^{(1,\dotsc,m)}}(\bm{z})- \nu_m^{\mathrm{RS}}(\bm{z}))\le\epsilon(\bm{z})\right\}.
\end{align*}
\begin{theorem}[Central limit theorem for the dense model]\label{thm:Gauss}
On the assumption of Theorem~\ref{thm:dense},
$\left\{\sqrt{N}\left(\nu_m^{\bm{x}}(\bm{z})-\nu_m^{\mathrm{RS}}(\bm{z})\right)\right\}_{\bm{z}\in\mathcal{X}^m}$
 weakly converges to the degenerate Gaussian distribution of zero mean and
the covariance matrix $(S'-S)(I_{|\mathcal{X}^m|}-JD^2gJ^t(S'-S))^{-1}$.
\end{theorem}
Let the overlaps $q_{ab}^{\bm{x}}:=\langle z^{(a)}z^{(b)}\rangle_{\nu_m^{\bm{x}}}$.
As a consequence of Theorem~\ref{thm:Gauss},
$\{\sqrt{N}(q^{\bm{x}}_{ab}-q^{\mathrm{RS}})\}_{a\in\{1,\dotsc,m\},\,b\in\{a,\dotsc,m\}}$ weakly converges to the Gaussian distribution of zero mean and
the covariance matrix $(U'-U)\left(I_{m(m+1)/2} -D^2g(U'-U)\right)^{-1}$.
This result is known for SK model at high temperature $\beta<1$ rigorously (without replica method nor cavity method)~\cite{comets1995sherrington}
where the covariance matrix is $1/(1-\beta^2) I_{m(m-1)/2}$.
Obviously, a local limit theorem also holds although it is not explicitly stated here due to the lack of the space.

\begin{figure*}[!b]
\hrulefill
\begin{align}
\mathbb{E}[Z^n]
&= \sum_{\{v(\bm{x})\}_{\bm{x}\in\mathcal{X}^n}} \binom{N}{\{v(\bm{x})\}_{\bm{x}\in\mathcal{X}^n}} \exp\left\{N\sum_{\bm{x}\in\mathcal{X}^n}\frac{v(\bm{x})}{N}\sum_{a=1}^nf(x^{(a)})
 + Ng\left(\left\{\sum_{\bm{x}\in\mathcal{X}^n}\frac{v(\bm{x})}{N}x^{(a)}x^{(b)}\right\}\right)\right\}\nonumber\\
&\sim \sum_{\{v(\bm{x})\}_{\bm{x}\in\mathcal{X}^n}, \|v(\bm{x})-N\nu^*(\bm{x})\|\le N^{\alpha}} \binom{N}{\{v(\bm{x})\}_{\bm{x}\in\mathcal{X}^n}} \exp\left\{N\sum_{\bm{x}\in\mathcal{X}^n}\frac{v(\bm{x})}{N}\sum_{a=1}^nf(x^{(a)})
 + Ng\left(\left\{\sum_{\bm{x}\in\mathcal{X}^n}\frac{v(\bm{x})}{N}x^{(a)}x^{(b)}\right\}\right)\right\}\label{eq:p1}\\
&\sim \frac{\sqrt{2\pi N}}{\prod_{\bm{x}\in\mathcal{X}^n}\sqrt{2\pi N\nu^*(\bm{x})}} \exp\left\{N\mathcal{H}(\nu^*)+ N\sum_{\bm{x}\in\mathcal{X}^n}\nu^*(\bm{x})\sum_{a=1}^nf(x^{(a)}) + Ng\left(\left\{\sum_{\bm{x}\in\mathcal{X}^n}\nu^*(\bm{x})x^{(a)}x^{(b)}\right\}\right)\right\}\label{eq:p2}\\
&\quad\cdot
\sum_{\{v(\bm{x})\}_{\bm{x}\in\mathcal{X}^n}, \|v(\bm{x})-N\nu^*(\bm{x})\|\le N^{\alpha}}
\exp\left\{-\sum_{\bm{x}\in\mathcal{X}^n}(v(\bm{x})-N\nu^*(\bm{x}))\log \nu^*(\bm{x}) -\frac12\sum_{\bm{x}\in\mathcal{X}^n}\frac{(N\nu^*(\bm{x})-v(\bm{x}))^2}{N\nu^*(\bm{x})}\right\}\nonumber\\
&\quad\cdot \exp\left\{N\sum_{\bm{x}\in\mathcal{X}^n}\frac{v(\bm{x})-N\nu^*(\bm{x})}{N}\sum_{a=1}^nf(x^{(a)})
 + Ng\left(\left\{\sum_{\bm{x}\in\mathcal{X}^n}\frac{v(\bm{x})}{N}x^{(a)}x^{(b)}\right\}\right) - Ng\left(\left\{\sum_{\bm{x}\in\mathcal{X}^n}\nu^*(\bm{x})x^{(a)}x^{(b)}\right\}\right)\right\}\nonumber\\
&\sim \frac{\sqrt{2\pi N}}{\prod_{\bm{x}\in\mathcal{X}^n}\sqrt{2\pi N\nu^*(\bm{x})}}
 \exp\left\{N\mathcal{H}(\nu^*)+N\sum_{\bm{x}\in\mathcal{X}^n}\nu^*(\bm{x})\sum_{a=1}^nf(x^{(a)}) + Ng\left(\left\{\sum_{\bm{x}\in\mathcal{X}^n}\nu^*(\bm{x})x^{(a)}x^{(b)}\right\}\right)\right\}\label{eq:p3}\\
&\quad\cdot
\sum_{\{v(\bm{x})\}_{\bm{x}\in\mathcal{X}^n}, \|v(\bm{x})-N\nu^*(\bm{x})\|\le N^{\alpha}}
 \exp\left\{-\sum_{\bm{x}\in\mathcal{X}^n}(v(\bm{x})-N\nu^*(\bm{x}))\log \nu^*(\bm{x}) -\frac12\sum_{\bm{x}\in\mathcal{X}^n}\frac{(N\nu^*(\bm{x})-v(\bm{x}))^2}{N\nu^*(\bm{x})}\right\}\nonumber\\
&\quad\cdot\exp\left\{N\sum_{\bm{x}\in\mathcal{X}^n}\frac{v(\bm{x})-N\nu^*(\bm{x})}{N}\sum_{a=1}^nf(x^{(a)})
+N \left[\sum_{\bm{x}\in\mathcal{X}^n}\frac{v(\bm{x})-N\nu^*(\bm{x})}{N}x^{(a)}x^{(b)} \right]^tD g\left(\left\{\sum_{\bm{x}\in\mathcal{X}^n}\nu^*(\bm{x})x^{(a)}x^{(b)}\right\}\right)\right\}\nonumber\\
&\quad\cdot \exp\left\{N \frac12 \left[\sum_{\bm{x}\in\mathcal{X}^n}\frac{v(\bm{x})-N\nu^*(\bm{x})}{N}x^{(a)}x^{(b)} \right]^t D^2 g\left(\left\{\sum_{\bm{x}\in\mathcal{X}^n}\nu^*(\bm{x})x^{(a)}x^{(b)}\right\}\right)
\left[\sum_{\bm{x}\in\mathcal{X}^n}\frac{v(\bm{x})-N\nu^*(\bm{x})}{N}x^{(a)}x^{(b)} \right]\right\}\nonumber\\
&\sim \frac{\sqrt{2\pi N}}{\prod_{\bm{x}\in\mathcal{X}^n}\sqrt{2\pi N\nu^*(\bm{x})}} \exp\left\{N\mathcal{H}(\nu^*)+N\sum_{\bm{x}\in\mathcal{X}^n}\nu^*(\bm{x})\sum_{a=1}^nf(x^{(a)}) + Ng\left(\left\{\sum_{\bm{x}\in\mathcal{X}^n}\nu^*(\bm{x})x^{(a)}x^{(b)}\right\}\right)\right\}\label{eq:p4}\\
&\quad\cdot
\sum_{\{\epsilon(\bm{x})\}_{\bm{x}\in\mathcal{X}^n},\|\epsilon(\bm{x})\|\le N^\alpha} \exp\left\{-\frac12\sum_{\bm{x}\in\mathcal{X}^n}\frac{\epsilon(\bm{x})^2}{N\nu^*(\bm{x})}
+ N \frac12 \left[\sum_{\bm{x}\in\mathcal{X}^n}\frac{\epsilon(\bm{x})}{N}x^{(a)}x^{(b)} \right]^t D^2 g
\left(\left\{\sum_{\bm{x}\in\mathcal{X}^n}\nu^*(\bm{x})x^{(a)}x^{(b)}\right\}\right)
\left[\sum_{\bm{x}\in\mathcal{X}^n}\frac{\epsilon(\bm{x})}{N}x^{(a)}x^{(b)} \right]\right\}\nonumber\\
&\sim \frac{\sqrt{2\pi N}}{\prod_{\bm{x}\in\mathcal{X}^n}\sqrt{2\pi N\nu^*(\bm{x})}} \exp\left\{N\mathcal{H}(\nu^*)+N\sum_{\bm{x}\in\mathcal{X}^n}\nu^*(\bm{x})\sum_{a=1}^nf(x^{(a)}) + Ng\left(\left\{\sum_{\bm{x}\in\mathcal{X}^n}\nu^*(\bm{x})x^{(a)}x^{(b)}\right\}\right)\right\}\label{eq:p5}\\
&\quad\cdot
N^{(|\mathcal{X}|^n-1)/2}\int \mathrm{d}\epsilon(\bm{x})\exp\left\{-\frac12\sum_{\bm{x}\in\mathcal{X}^n}\frac{\epsilon(\bm{x})^2}{\nu^*(\bm{x})}
+ \frac12 \left[\sum_{\bm{x}\in\mathcal{X}^n}\epsilon(\bm{x})x^{(a)}x^{(b)} \right]^t D^2 g\left(\left\{\sum_{\bm{x}\in\mathcal{X}^n}\nu^*(\bm{x})x^{(a)}x^{(b)}\right\}\right)
\left[\sum_{\bm{x}\in\mathcal{X}^n}\epsilon(\bm{x})x^{(a)}x^{(b)} \right]\right\}\nonumber\\
&= \exp\left\{N\mathcal{H}(\nu^*)+N\sum_{\bm{x}\in\mathcal{X}^n}\nu^*(\bm{x})\sum_{a=1}^nf(x^{(a)}) + Ng\left(\left\{\sum_{\bm{x}\in\mathcal{X}^n}\nu^*(\bm{x})x^{(a)}x^{(b)}\right\}\right)\right\}
\frac1{\prod_{\bm{x}\in\mathcal{X}^n} \sqrt{\nu^*(\bm{x})}}\det\left(H^t(B-JD^2gJ^t)H\right)^{-\frac12}\nonumber\\
&= \exp\left\{N\mathcal{H}(\nu^*)+N\sum_{\bm{x}\in\mathcal{X}^n}\nu^*(\bm{x})\sum_{a=1}^nf(x^{(a)}) + Ng\left(\left\{\sum_{\bm{x}\in\mathcal{X}^n}\nu^*(\bm{x})x^{(a)}x^{(b)}\right\}\right)\right\}
\det\left(I_{|\mathcal{X}|^n-1} -H^tJ(D^2g)J^tH(H^tBH)^{-1}\right)^{-\frac12}\nonumber
\end{align}
\end{figure*}

\section{Random sparse regular factor graph ensembles}
In this section, we deal with the random regular factor graph ensembles.
The calculation of the exponent of the partition function using the method of types is proposed in~\cite{mori2011connection}
while the basic idea of the type of factor graph is mentioned in~\cite{vontobel2010counting}.
In this section, similarly to the previous section, the central approximation is used for deriving the constant factor.

A factor graph is a bipartite graph consisting of variable nodes and factor nodes, defining a probability distribution
\begin{align*}
p(\bm{x}) &:= \frac1Z \prod_a f(\bm{x}_{\partial a}),&
Z &:= \sum_{\bm{x}\in\mathcal{X}^N}\prod_a f(\bm{x}_{\partial a})
\end{align*}
where $a$ is the index of the factor nodes and where $\partial a$ is the set of indices of variable nodes connected to the factor node $a$.
Let $l$ and $r$ be degrees of variable and factor nodes of regular factor graph ensembles, respectively.
The random connection of edges is chosen uniformly from the $(Nl)!$ possible connections.
Let $\mathbb{E}[\cdot]$ denote the expectation on random connection of edges.
Let \textit{variable-type} $v$ denote the type of variable nodes, i.e., there exists $v(x)$ variable nodes of value $x\in\mathcal{X}$.
Let \textit{factor-type} $u$ denote the type of factor nodes, in which the value of a factor node is regarded as
the values of variable nodes connected to the factor node,
i.e., there exists $u(\bm{x})$ factor nodes connecting variable nodes of values $x_1,x_2,\dotsc,x_r$.
Here, the order of values is distinguished for general $f(\bm{x})$ which is not invariant under permutations of the arguments $\bm{x}\in\mathcal{X}^r$.
Let $N(v,u)$ be the number of assignments with variable-type $v$ and factor-type $u$.
The partition function $Z$ is then given in terms of types as 
\begin{equation*}
Z =\sum_{v,u} N(v,u)\prod_{\bm{x}\in\mathcal{X}^r}f(\bm{x})^{u({\bm{x}})}.
\end{equation*}
In the summation above, the types $v$ and $u$ have to satisfy the condition for consistency
\begin{equation}
\sum_{\bm{x}\in\mathcal{X}^r} N_z(\bm{x})u(\bm{x}) = lv(z)
\label{eq:cond}
\end{equation}
where $N_z(\bm{x})$ denotes the number of $z\in\mathcal{X}$ in $\bm{x}\in\mathcal{X}^r$.
The expected number $N(v,u)$ of assignments with variable-type $v$ and factor-type $u$ is
\begin{equation*}
\mathbb{E}[N(v,u)]
=
\binom{N}{\{v(x)\}_{x\in\mathcal{X}}}
\binom{\frac{l}{r}N}{\{u(\bm{x})\}_{\bm{x}\in\mathcal{X}^r}}
\frac{\prod_{x\in\mathcal{X}}
(v(x)l)!}{(Nl)!}.
\end{equation*}
One thus obtains the exponent as 
\begin{align}
&F:=\lim_{N\to\infty} \frac1N \log \mathbb{E}[Z]\nonumber\\
&=\max_{\nu,\mu}\bigg\{
\frac{l}{r}\mathcal{H}(\mu) - (l-1)\mathcal{H}(\nu)
+\frac{l}{r}\sum_{\bm{x}\in\mathcal{X}^r} \mu(\bm{x})\log f(\bm{x})\bigg\}
\label{eq:max}
\end{align}
where $\nu$ and $\mu$ are probability measures on $\mathcal{X}$ and $\mathcal{X}^r$, respectively, satisfying
\begin{equation*}
\frac1{r}\sum_{i=1}^r\sum_{\substack{\bm{x}\setminus x_i\\ x_i=z}}\mu(\bm{x}) = \nu(z),\hspace{2em}\forall z\in\mathcal{X}.
\end{equation*}
The above maximization problem can be regarded as the minimization problem of the Bethe free energy on the averaged model~\cite{mori2011connection}.

For obtaining the constant factor, the central approximation is used similarly as in the previous section.
The derivation is omitted for the lack of space.
The unique difference is that the condition~\eqref{eq:cond} affects the step size in the Riemann integral formula.
By leaving the product of step sizes as the unknown variable $s$, the following theorem is obtained.
\begin{theorem}[Central approximation for random regular factor graph ensembles]\label{thm:rg}
Assume that the solution of the maximization problem~\eqref{eq:max} is unique and is denoted by $(\nu^*(\bm{x}),\mu^*(\bm{x}))$.
Furthermore, assume $\nu^*(x)>0$ for all $x\in\mathcal{X}$, and 
\begin{equation*}
\det\left(I_{|\mathcal{X}|}-C(V'-V)\right)>0
\end{equation*}
where $C$, $V'$ and $V$ are $|\mathcal{X}|\times |\mathcal{X}|$ matrices defined by
\begin{align*}
C(x,x')&=\begin{cases}
\frac{r(l-1)}{l\nu^*(x)},&\text{if } x=x'\\
0,&\text{if } x \ne x'
\end{cases}\\
V'(x,x') &= \frac1{r^2}\sum_{k,k'=1}^r\sum_{\substack{\bm{x}\in\mathcal{X}^r,\\ x_k=x, x_{k'}=x'}}\mu^*(\bm{x})\\
V(x,x') &= \nu^*(x)\nu^*(x').
\end{align*}
Then,
\begin{equation*}
\mathbb{E}[Z]= 
\mathrm{e}^{NF}l^{\frac{|\mathcal{X}|-1}2}\frac1{s}\det\left(I_{|\mathcal{X}|}-C(V'-V)\right)^{-\frac12}
(1+o(1))
\end{equation*}
where $F$ is given by~\eqref{eq:max},
and where $s$ is some integer depending on $l$, $r$ and the support of $f(\bm{x})$.
\end{theorem}

Let $\mathcal{S}\subseteq\mathcal{X}^r$ be the support of $f(\bm{x})$.
In order to obtain $s$, the following condition for $\{\epsilon(\bm{x})\}_{\bm{x}\in\mathcal{S}}$, which play the same role in the analysis as those in~\eqref{eq:p4}, must be considered:
\begin{equation}\label{eq:icond}
\prod_{x\in\mathcal{X}\setminus 0}\mathbb{I}\left\{\sum_{\bm{x}\in\mathcal{S}\setminus\bm{x_0}}\left(N_x(\bm{x})-N_x(\bm{x_0})\right) \epsilon(\bm{x})
 \text{ is a multiple of } l\right\}
\end{equation}
where 0 and $\bm{x}_0$ are any fixed elements in $\mathcal{X}$ and $\mathcal{S}$, respectively.
Although we have not obtained a general result about $s$,
there are several cases where $s$ can easily be specified.
When $l$ is a prime, \eqref{eq:icond} defines simultaneous linear equations on the finite field $\mathbb{F}_l$.
Hence, $s=l^{c}$ where $c$ denotes the rank of the simultaneous linear equations.
When $\mathcal{X}=\{0,1\}$, one has 
$s = l/g$ where $g:=\gcd(\{N_0(\bm{x})-N_0(\bm{x}_0)\}_{\bm{x}\in\mathcal{S}\setminus \bm{x}_0},l)$.
As a consequence, the asymptotic expected number of codewords of LDPC codes is obtained up to the constant factor~\cite{1715529}.

The annealed version of Theorem~\ref{thm:Gauss} for random regular factor graph ensembles is obtained as follows.
\begin{theorem}[Central limit theorem for random regular factor graph ensembles]~\label{thm:annealG}
On the assumption of Theorem~\ref{thm:rg},
\begin{align*}
&\lim_{N\to\infty}\frac{\mathbb{E}\left[\sum_{\bm{x}\in\mathcal{X}^N} \prod_a f(\bm{x}_{\partial a})
\prod_{\bm{z}\in\mathcal{S}}\mathbb{I}\left\{\sqrt{N}\left(\frac{u(\bm{z})}{N}-\mu^*(\bm{z})\right)\le t(\bm{z})\right\}\right]}{\mathbb{E}[Z]}\\
&=\Pr\left(\cap_{\bm{z}\in\mathcal{S}}X_{\bm{z}}\le t(\bm{z})\right)
\end{align*}
where $\{X_{\bm{z}}\}_{\bm{z}\in\mathcal{S}}$ is the degenerate Gaussian distribution of zero mean and
the covariance matrix $(T'-T)(I_{|\mathcal{X}^r|}- KCK^t(T'-T))^{-1}$
where $T'$ is an $|\mathcal{X}^r|\times|\mathcal{X}^r|$ diagonal matrix defined by $T'(\bm{x},\bm{x})=\mu^*(\bm{x})$,
where $T$ is an $|\mathcal{X}^r|\times|\mathcal{X}^r|$ matrix defined by $T(\bm{x},\bm{x}')=\mu^*(\bm{x})\mu^*(\bm{x}')$,
and where $K$ is an $|\mathcal{X}^r|\times|\mathcal{X}|$ matrix defined by $K(\bm{x},x)=N_x(\bm{x})/r$.
\end{theorem}
For the type of variable nodes $\{v(x)\}_{x\in\mathcal{X}}$, a similar result is obtained with the covariance matrix 
$(V'-V)(I_{|\mathcal{X}|}-C(V'-V))^{-1}$.
As mentioned in the previous section, a local limit theorem also holds.
The results in this section can be generalized to the quenched version by using the replica method similarly to the previous section.

\section*{Acknowledgment}
{\normalsize The work of RM was supported by the Grant-in-Aid for Scientific Research for JSPS Fellows (22$\cdot$5936), JSPS, Japan.}

\bibliographystyle{IEEEtran}
\bibliography{IEEEabrv,ldpc}

\end{document}